%% file: lam_pol_paper.tex
\documentclass[a4paper,12pt]{article}
\usepackage{
amssymb,amsfonts,amsmath,epsfig,
mathrsfs,fancyhdr,subfigure,graphics,
graphicx,array,textcomp,afterpage}
\newcommand{\xout}[1]{}

\newcommand{\hb} {\mbox{\sffamily HERA\protect\rule[.5ex]{1.ex}{.11ex}B}\ }

\newcommand{\lam}{$\Lambda \thickspace$}

\newcommand{\ks}{$K^{0}_{S} \thickspace$}
\newcommand{\alam}{$\overline{\Lambda} \thickspace$}
\newcommand{\alamkomma}{$\overline{\Lambda}, \thickspace$}

\newcommand{\ptpunktum}{$p_{\bot}.  \thickspace$}

\newcommand{\xfkomma}{$x_F, \thickspace$}
\newcommand{\xfpunktum}{$x_F. \thickspace$}
\newcommand{\pt}{$p_{\bot} \thickspace$}
\newcommand{\xf}{$x_F \thickspace$}

\newcommand{\lamalam}{$\Lambda/\overline{\Lambda} \thickspace $}
\newcommand{\lamalams}{$\Lambda/\overline{\Lambda}$'s \thickspace }
\newcommand{\lamalamskomma}{$\Lambda/\overline{\Lambda}$'s, \thickspace }
\newcommand{\lamalamm}{$\Lambda/\overline{\Lambda}$) \thickspace\negthinspace }

\newcommand{\lams}{$\Lambda$'s\thickspace}
\newcommand{\lamskomma}{$\Lambda$'s,\thickspace}

\begin{document}
\textbf{Polarization of $\mathbf{\Lambda}$ and $\mathbf{\overline{\Lambda}}$ in 920\,GeV Fixed-Target Proton-Nucleus Collisions}\\
\begin{flushleft} 
\input{authors_epjc.tex}
\end{flushleft}

\newpage
\textbf{
Abstract}\\

\emph{A measurement of the polarization of \lam and \alam baryons produced in
$pC$ and $pW$ collisions at} $\sqrt{s}=41.6$~GeV \emph{has been performed with the \hb
spectrometer. The measurements cover the kinematic range of} $0.6$~GeV/c$~<p_{\bot}<1.2$~GeV/c \emph{in transverse momentum and} $-0.15<x_F<0.01$ \emph{in
Feynman-x. The polarization results from the two different targets agree within the statistical error. In the combined data set, the largest deviation from zero,~$+0.054 \pm 0.029$, is measured for $x_F\lesssim -0.07$. Zero polarization is expected at $x_F=0$ in the absence of nuclear effects.
The polarization results for the \lam agree with a parametrization of
previous measurements which were performed at positive \xf values, 
where the \lam polarization is negative.
Results of \alam polarization measurements are consistent with zero.}

\section{Introduction}
\label{sec:intro}

Previous measurements (see
e.g.~Refs.~\cite{Heller:1978ty,Ramberg:1994tk}) have, contrary to
expectations, shown that \lams and other hyperons produced in unpolarized hadron-hadron
interactions are transversely polarized. For \lamskomma the magnitude of the
polarization is observed to depend on the kinematic variables. For fixed-target $pA$ interactions usually the \lam momentum
transverse to the beam direction, $p_{\bot}$, and its Feynman-x,
taken to be $x_F={2p_{\ell}}/{\sqrt{s}}$, are used. Here $p_{\ell}$ is the longitudinal momentum of
the hyperon relative to beam direction as measured in the center of
mass of the beam proton and target nucleon. The
magnitude of the polarization is observed to increase with \pt and decrease as
$\left|x_F\right|$ approaches zero.  No existing model adequately
describes the observations. (For general introductions to the topic of
\lam polarization and overviews of previous results and models see
Refs.~\cite{Heller:1996pq,Soffer:1999ww}.)  Additional experimental
input in previously unmeasured kinematic regions could provide additional insight
into the mechanism responsible for the polarization.

Most previous measurements were performed at positive $x_F$, the only
exceptions being low statistics measurements from bubble chamber
experiments \cite{Yuldashev:1991az} which probe the polarization over
the full phase space.  In this letter, we report a new measurement of
\lam and \alam (henceforth designated \lamalamm polarizations in
inclusive 920~GeV proton-nucleus interactions, predominately at
negative $x_F$ and in the \pt range of $0.6$~GeV/c
$<p_{\bot}<1.2$~GeV/c.  

The \lam polarization is inferred from the magnitude of
the angular asymmetry of protons resulting from the decay
$\Lambda\rightarrow p \pi^-$, as observed in the \lam rest-frame. For
each event the coordinate system is defined such that the $\vec{n}_z$
axis coincides with the boost vector from the laboratory system to the
\lam rest-frame. The $\vec{n}_x$ direction is the normal to the
production plane as defined by the cross product of the beam direction
as seen in the \lam rest frame with the $\vec{n}_z$ axis
($\vec{p}_{beam} \times \vec{n}_z$) and
$\vec{n}_y=\vec{n}_{z}\times\vec{n}_{x}$. Since the \lams are produced
via parity-conserving strong interactions, polarization can only occur
transverse to the production plane \cite{Perkins:1982xb},
corresponding to the $\vec{n}_x$ direction. The polarization is measurable since the \lams decay via a parity-nonconserving weak process.

The expected intensities are:

\begin{equation}
\begin{split}
\frac{dN}{d\cos{\theta_x}}&\propto A(\cos{\theta_x})(1+\alpha_{\Lambda} P_{\Lambda}\cos{\theta_x})\\
\frac{dN}{d\cos{\theta_y}}&\propto A(\cos{\theta_y})\\
\frac{dN}{d\cos{\theta_z}}&\propto A(\cos{\theta_z})
\end{split}
\label{eq:intxyz}
\end{equation}

where $\cos{\theta_i}=\vec{n}_i \cdot \vec{n}_{proton }\thickspace
\text{for} \thickspace i=x,y,z$. $P_{\Lambda}$ is the 
polarization, $A$ is the detector acceptance and $\alpha_{\Lambda}$ is
the asymmetry parameter of the \lam decay. For the \alamkomma the
equations are modified by substituting: $\vec{n}_{proton}\rightarrow
\vec{n}_{anti-proton }$ and
$\alpha_{\Lambda}\rightarrow\alpha_{\overline{\Lambda}}=-\alpha_{\Lambda}$.

\section{Detector, Data Sample and Event Selection}

The data sample used for this analysis was collected with the
fixed-target \hb spectro-meter operating at the 920~GeV proton storage
ring of HERA, at DESY.  The target consists of thin wires of various
materials -- for this measurement, carbon (C) and tungsten (W) --
dynamically positioned in the halo of the proton beam.  Particles
produced in collisions are measured using a variety of sub-detector
systems, the most important for this paper being the silicon vertex
detector (VDS)~\cite{Bauer:2003} and the outer tracker
(OTR)~\cite{Albrecht:2005}.  The VDS is positioned immediately
downstream of the target and consists of 8 planar stations with a
total of 64 double-sided silicon microstrip detectors. 
The VDS is followed by a large aperture 2.13 T$\cdot$m magnet
and the OTR, which consists of 7 planar stations of honeycomb drift
chambers.  The spectrometer has a large angular coverage: 15~mrad to
220~mrad in the horizontal (bending) plane and 15~mrad to 160~mrad
in the vertical plane. A Ring Imaging Cerenkov detector (RICH) and an
electromagnetic calorimeter (ECAL) cover the full aperture and, for
the purposes of this measurement, are only used to provide a minimum
bias trigger: events are required to have either 20 hits in the RICH
(corresponding to 60\% of the expected yield of a single relativistic
charged particle) or at least one ECAL cluster with an energy of at
least 1~GeV. More details on the spectrometer can be found
in Ref.~\cite{bbbar:2005} and references therein.

The data sample consists of a total of 119~million events from two
targets:  55~million from the carbon target sample and
  64~million events from the tungsten target sample (henceforth
referred to as the C-target and W-target samples). The events are
selected from single wire runs and consist
mainly of single interactions, with approximately 10\% having more
than one interaction.

The detector acceptance is determined from Monte Carlo simulations
(MC). \textsc{Fritiof 7.02}~\cite{Pi:1992ug} is used as event generator,
and a {\sc GEANT 3.21}-based detector model simulates the detector
response~\cite{GEANT}. The generated decay angle
distributions are flat in $cos(\theta_i)$.  The generated \pt and \xf
distributions of the \lamalams were tuned such that the reconstructed
MC distributions are in agreement with the uncorrected data in the
kinematic range of the measurement.

Segments of tracks are reconstructed in the VDS and OTR independently requiring at least five and six hits, respectively. Segments are then combined to tracks with the constraint that each segment is only allowed in one track.
No particle identification cuts are
applied.  A \lam candidate is initially identified as two
oppositely-charged tracks forming a common vertex downstream of the target.  
Using the signal, $S$, and background, $B$, from data, an optimization of the signal significance, 
$S/\sqrt{S+B}$, is performed with respect to three discriminating variables, with the resulting cuts: 
\begin{itemize}
\item The impact parameter of the \lam candidate to the closest primary vertex is required to be less than $0.063~\mathrm{cm}$.  
\item The maximal allowed distance of closest approach between the two decay tracks is $0.15~\mathrm{cm}$.  
\item The flight path of the \lam candidate times 
the sum of the momenta of the decay tracks transverse to the direction 
of propagation 
of the \lam candidate is required to be larger than 0.15~cm$\cdot$GeV/c.
\end{itemize}

\begin{centering}
\begin{figure}
\begin{minipage}{\linewidth}
\centering
  \epsfig{figure=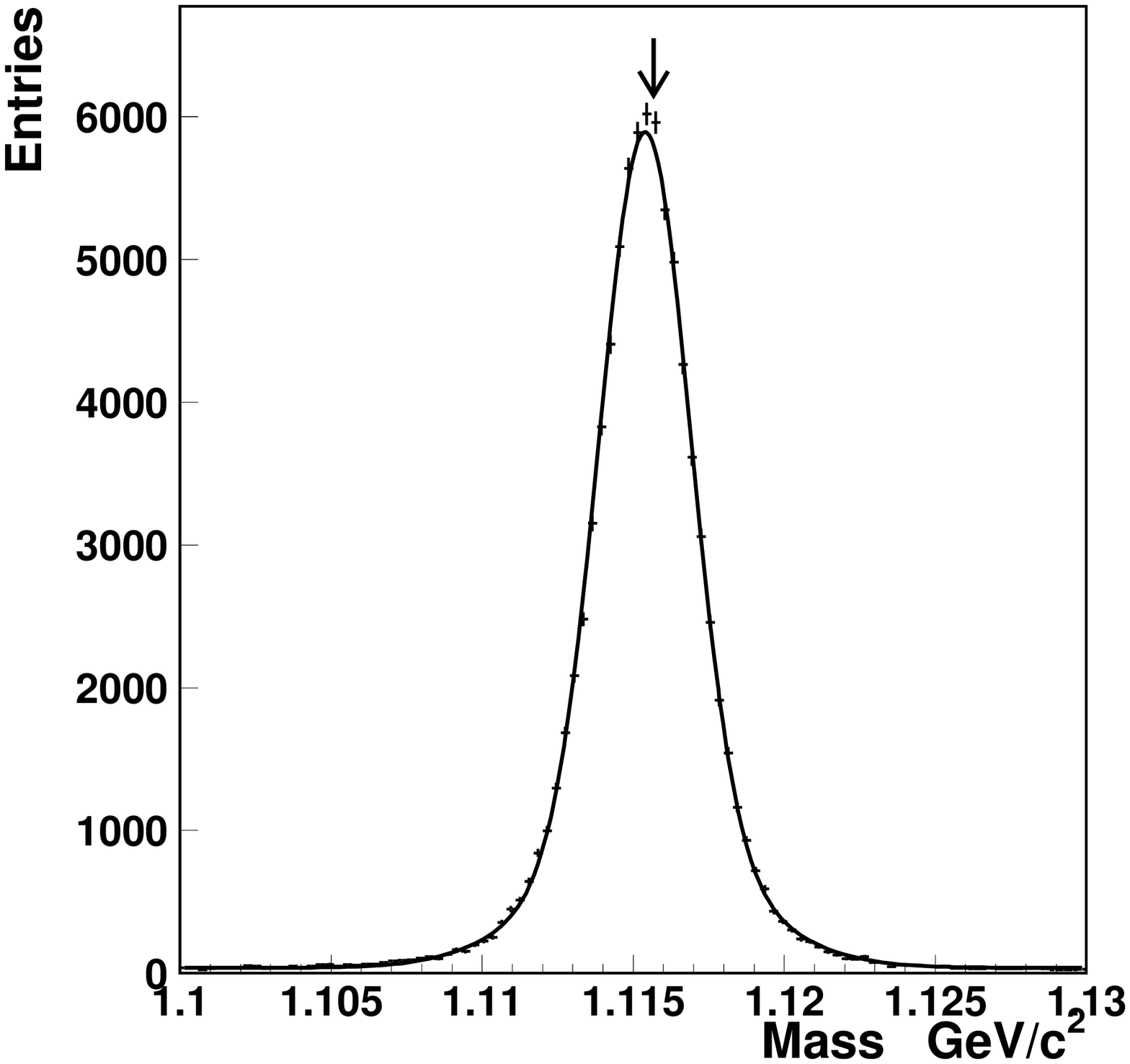,width=0.48\linewidth}
  \epsfig{figure=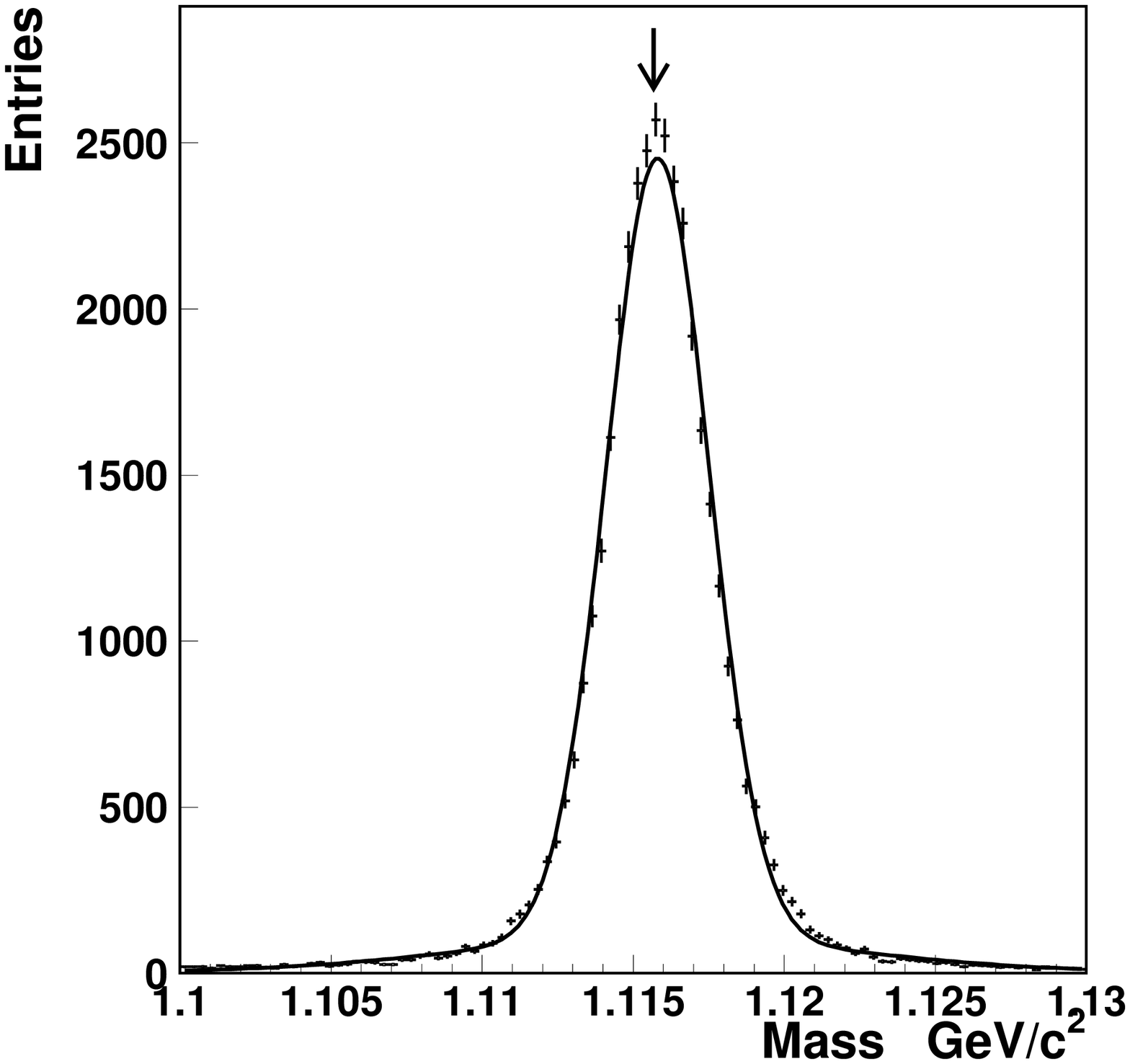,width=0.48\linewidth}
\begin{picture}(0,0)
\put(-363,100){$\Lambda$}
\put(-425,187){\Large{(a)}}
\put(-332,192){{\textbf{PDG value}}}
\put(-193,187){\Large{(b)}}
\put(-100,192){{\textbf{PDG value}}}
\put(-130,100){$\overline{\Lambda}$}
\end{picture}
\end{minipage}\hfill
\caption{(a) $p\pi^-$ invariant mass distribution for selected \lam candidates,
and, (b) $\overline{p}\pi^+$ invariant mass distribution for selected \alam candidates
for the W-target sample.}
\label{fig:masspeak}
\end{figure}
\end{centering}

An effect of the first cut is, according to MC studies, that the fraction
of cascade \lams is suppressed by a factor of approximately 10, so that only
 $\approx 1\%$ of \lams in the final signal originates from cascade decays.
The invariant mass of the
\lam candidate is calculated under the assumption that the positive
track is a proton and the negative track is a pion. Only candidates 
with masses in the mass range: [1.10; 1.13]~GeV/c$^2$ are considered 
in the analysis. \lam candidates
which are also consistent with a \ks hypothesis are rejected by eliminating those candidates whose invariant mass when calculated under the assumption that both tracks are pions, lies within the \ks mass window (i.e. less than $15~\mathrm{MeV/c^2}$ from the nominal \ks mass). This cut ensures that
the percentage of misidentified \lams is less than 0.3\%. The
analogous search is made for $\overline{\Lambda}$'s. Finally, we consider only \lamalams in
the range $0.6~\mathrm{GeV/c} < p_{\bot}  < 1.2~\mathrm{GeV/c}$ and $-0.15 < x_F < 0.01$. 
The invariant mass distributions
for selected $p\pi^-$ and $\overline{p}\pi^+$ candidates from the
W-target sample are shown in Fig.~\ref{fig:masspeak}(a) and
Fig.~\ref{fig:masspeak}(b), respectively.  The distributions for the
C-target sample are similar. According to fits using
two Gaussians to describe the signal and a second order polynomial to 
describe the background, the C-target 
and W-target samples contain (47K/23K) and (84K/37K) \lamalamskomma respectively.
Depending on the kinematic range (see Sect.~\ref{sec:results}), 
the background constitutes $\sim 4\%$ of the signal, but since events from 
the side-bins of the \lam mass distribution show no significant dependence 
on $cos(\theta _x)$, the background contribution to the polarization is 
negligible. 

\section{Results}
\label{sec:results}

\begin{figure}%[h!]
\begin{minipage}{\linewidth}
  \centering
  \epsfig{figure=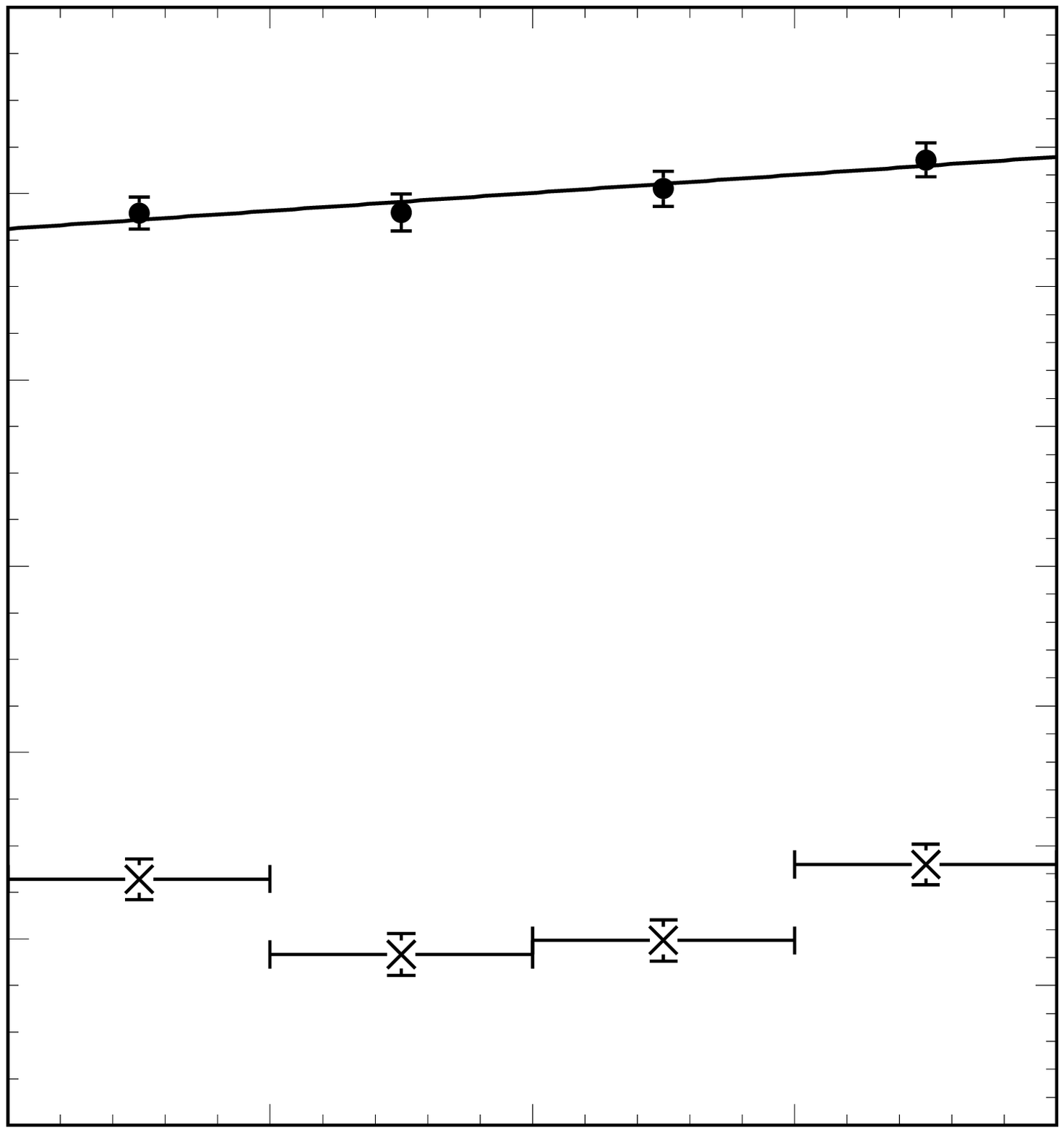,width=0.3\linewidth}
  \quad\medspace\thinspace\thinspace
  \epsfig{figure=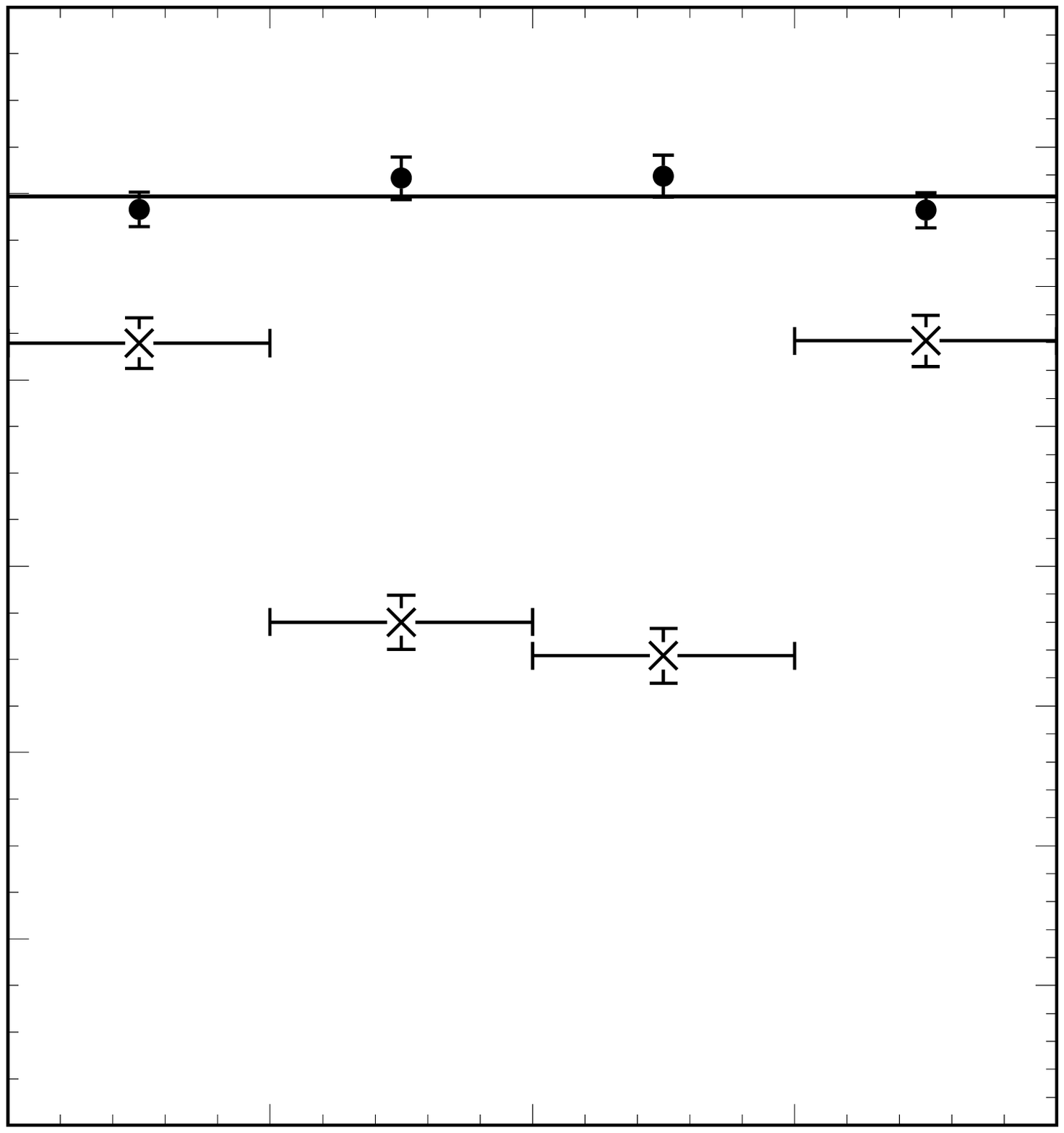,width=0.3\linewidth}
 \quad \medspace\thinspace\thinspace
  \epsfig{figure=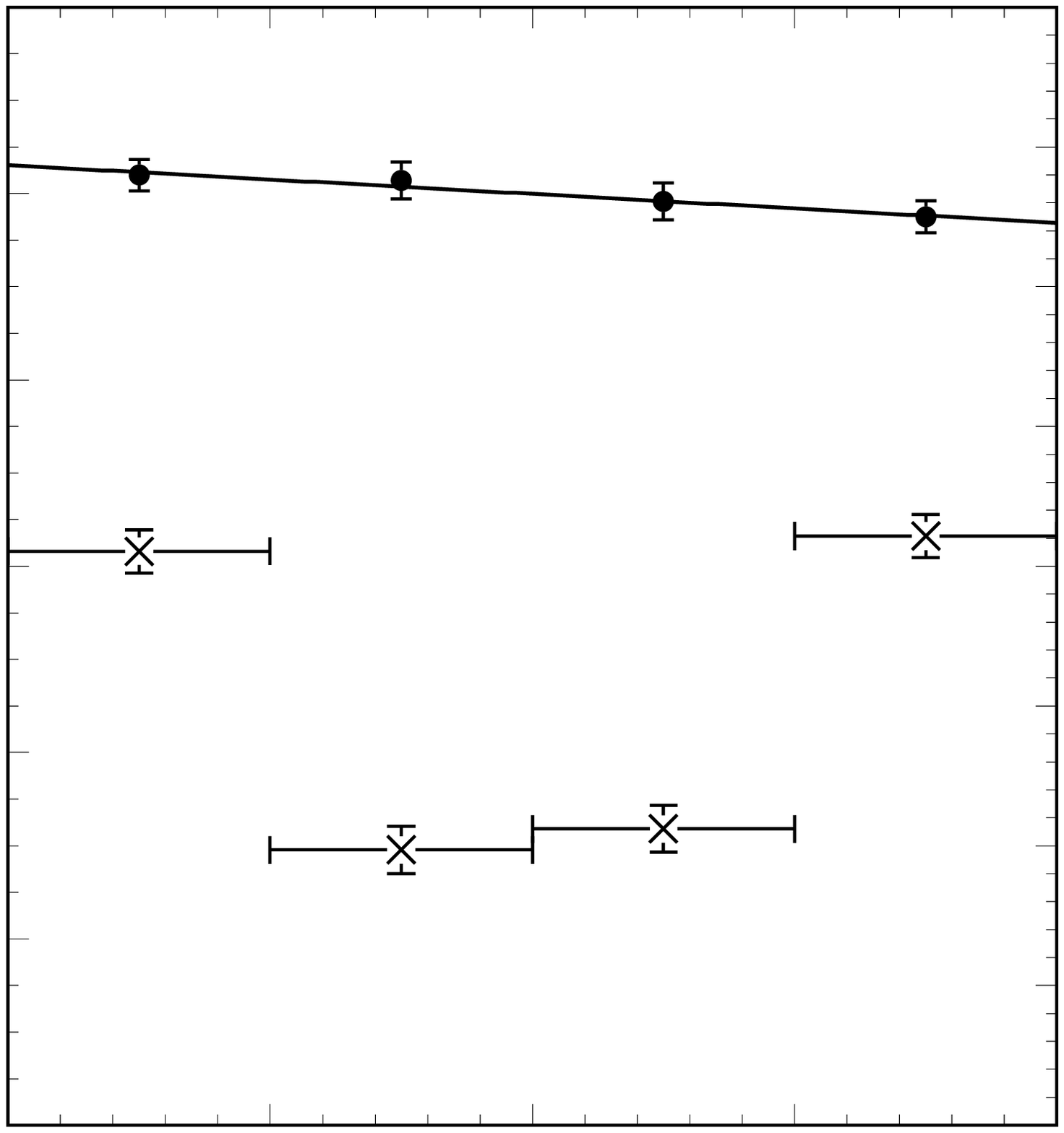,width=0.3\linewidth}
\begin{picture}(0,0)
\put(-451,140){$-0.15<x_F<-0.07$}
\put(-286,140){$-0.07<x_F<-0.04$}
\put(-120,140){$-0.04<x_F<0.01$}
\put(-415,-3){$cos(\theta _x)$}
\put(-248,-3){$cos(\theta _x)$}
\put(-83,-3){$cos(\theta _x)$}

\put(-353,8){\scriptsize{1.0}}
\put(-463,8){\scriptsize{-1.0}}
\put(-406,8){\scriptsize{0.0}}
\put(-435,8){\scriptsize{-0.5}}
\put(-379,8){\scriptsize{0.5}}

\put(-467,14){\scriptsize{0.0}}
\put(-467,31){\scriptsize{0.2}}
\put(-467,50){\scriptsize{0.4}}
\put(-467,69){\scriptsize{0.6}}
\put(-467,88){\scriptsize{0.8}}
\put(-467,107){\scriptsize{1.0}}
\put(-467,126){\scriptsize{1.2}}

\put(-344,14){\scriptsize{0.4}}
\put(-344,27){\scriptsize{0.5}}
\put(-344,41){\scriptsize{0.6}}
\put(-344,55){\scriptsize{0.7}}
\put(-344,70){\scriptsize{0.8}}
\put(-344,84){\scriptsize{0.9}}
\put(-344,98){\scriptsize{1.0}}$\medspace$

\put(-485,10){\rotatebox{90.0}{\scriptsize{ $\frac{dN}{dcos(\theta_x)}|_{data}/\frac{dN}{dcos(\theta_x)}|_{MC}$ ($\bullet$)}}}
\put(-335,42){\rotatebox{90.0}{\scriptsize{efficiency ($\times$)}}}

\put(-191,8){\scriptsize{1.0}}
\put(-301,8){\scriptsize{-1.0}}
\put(-244,8){\scriptsize{0.0}}
\put(-274,8){\scriptsize{-0.5}}
\put(-217,8){\scriptsize{0.5}}

\put(-305,14){\scriptsize{0.0}}
\put(-305,31){\scriptsize{0.2}}
\put(-305,50){\scriptsize{0.4}}
\put(-305,69){\scriptsize{0.6}}
\put(-305,88){\scriptsize{0.8}}
\put(-305,107){\scriptsize{1.0}}
\put(-305,126){\scriptsize{1.2}}
\put(-182,14){\scriptsize{0.4}}
\put(-182,27){\scriptsize{0.5}}
\put(-182,41){\scriptsize{0.6}}
\put(-182,55){\scriptsize{0.7}}
\put(-182,70){\scriptsize{0.8}}
\put(-182,84){\scriptsize{0.9}}
\put(-182,98){\scriptsize{1.0}}$\medspace$
\put(-323,10){\rotatebox{90.0}{\scriptsize{ $\frac{dN}{dcos(\theta_x)}|_{data}/\frac{dN}{dcos(\theta_x)}|_{MC}$ ($\bullet$)}}}
\put(-173,42){\rotatebox{90.0}{\scriptsize{efficiency ($\times$)}}}
\put(-30,8){\scriptsize{1.0}}
\put(-140,8){\scriptsize{-1.0}}
\put(-83,8){\scriptsize{0.0}}
\put(-113,8){\scriptsize{-0.5}}
\put(-56,8){\scriptsize{0.5}}
\put(-144,14){\scriptsize{0.0}}
\put(-144,31){\scriptsize{0.2}}
\put(-144,50){\scriptsize{0.4}}
\put(-144,69){\scriptsize{0.6}}
\put(-144,88){\scriptsize{0.8}}
\put(-144,107){\scriptsize{1.0}}
\put(-144,126){\scriptsize{1.2}}
\put(-21,14){\scriptsize{0.4}}
\put(-21,27){\scriptsize{0.5}}
\put(-21,41){\scriptsize{0.6}}
\put(-21,55){\scriptsize{0.7}}
\put(-21,70){\scriptsize{0.8}}
\put(-21,84){\scriptsize{0.9}}
\put(-21,98){\scriptsize{1.0}}$\medspace$
\put(-162,10){\rotatebox{90.0}{\scriptsize{ $\frac{dN}{dcos(\theta_x)}|_{data}/\frac{dN}{dcos(\theta_x)}|_{MC}$ ($\bullet$)}}}
\put(-12,42){\rotatebox{90.0}{\scriptsize{efficiency ($\times$)}}}
\end{picture}
\end{minipage}
\caption{Corrected $cos(\theta _x)$ distributions ($\bullet$) and
  linear fit (left ordinate) and reconstruction efficiency ($\times$)
  for the given \xf and \pt range (right ordinate) for \lams of the
  W-target data. The low efficiency seen on the
  left plot is mainly due to the magnet which tends to bend the pion
  trajectories out of the detector acceptance for backward events.}
 \label{fig:cosxrat}
 \end{figure}

The polarization is determined separately in three \xf intervals of
similar event statistics:
$[-0.15;-0.07], [-0.07;-0.04]$ and $[-0.04;0.01]$.  For each \xf
interval, and for both real data and MC, the events are split into
four bins of equal size in $cos(\theta_x)$ and the $p\pi^-$ mass
spectra for each bin are fitted. The corrected $cos(\theta_x)$
distribution is the ratio: $\frac{dN}{dcos(\theta_x)}|_{data}/\frac{dN}{dcos(\theta_x)}|_{MC}$ of data to MC normalized to the same total
number of events. The resulting corrected distributions are plotted in
Fig.~\ref{fig:cosxrat}. Since the MC sample is generated flat in 
$cos(\theta_x)$, this ratio
should be flat if the \lamalams are unpolarized, and otherwise be a
linear function of $cos(\theta_x)$ according to Eq.~(\ref{eq:intxyz}).

The results are summarized in
Tables~\ref{tab:c} and~\ref{tab:w} for the C-target and W-target
samples separately.  The measured \lamalam polarizations in C-target and
W-target samples are consistent in all \xf intervals. We therefore
follow the approach of previous measurements and present results
averaged over the two data samples, see Table~\ref{tab:combined}.
The statistical uncertainties of the \alam polarization measurements
are larger than the corresponding \lam measurements due to the factor
of approximately two difference in statistics.

\begin{table}[!htbp]
  \centering
\setlength{\extrarowheight}{2.0pt}
  \begin{tabular}[!ht]{    >{ $}l<{$  }>{ $}c<{$}> {$}c<{$  }>{ $}c<{$}    }
  \mathbf{\Lambda}  &  <p_{\bot}> & <x_F> & Polarization  \\
&   \text{[GeV/c]}& &  \\ \hline
&  0.82&-0.099&\quad\negthinspace\negthinspace 0.046 \pm 0.031(stat) \\
&  0.81&-0.054&\quad\negthinspace\negthinspace 0.017 \pm 0.031(stat) \\
&  0.84&-0.020&-0.018 \pm 0.026(stat)\\   \hline
\mathbf{\overline{\Lambda}} \\ \hline
&  0.81&-0.097&-0.037  \pm0.051(stat) \\
&  0.80&-0.054&\quad\negthinspace\negthinspace 0.032 \pm0.051(stat) \\
&  0.83&-0.020&-0.035\pm0.044(stat)\\  \hline
  \end{tabular}
  \caption[Comparison of measurements obtained in this study with previous measurements]{\lamalam polarization results for the C-target sample in three bins of \xfpunktum The average \pt and \xf for each bin are also given.}
  \label{tab:c}
\end{table}
\begin{table}[!htbp]
  \centering
\setlength{\extrarowheight}{2.0pt}
  \begin{tabular}[!ht]{    >{ $}l<{$  }>{ $}c<{$}> {$}c<{$  }>{ $}c<{$}    }
  \mathbf{\Lambda}  &  <p_{\bot}> & <x_F> & Polarization  \\
&   \text{[GeV/c]}& &  \\ \hline
&  0.83&-0.099&\quad\negthinspace\negthinspace 0.060 \pm 0.025(stat)\\
&  0.82&-0.055&\quad\negthinspace\negthinspace 0.000 \pm 0.027(stat)\\
&  0.84&-0.020&-0.048 \pm 0.024(stat)\\   \hline
\mathbf{\overline{\Lambda}} \\ \hline
&  0.82&-0.097&-0.017  \pm0.037(stat)\\
&  0.82&-0.054&\quad\negthinspace\negthinspace 0.026 \pm0.036(stat) \\
&  0.83&-0.020&-0.019\pm0.030(stat)\\  \hline
  \end{tabular}
  \caption[Comparison of measurements obtained in this study with previous measurements]{\lamalam polarization results for the W-target sample in three bins of \xfpunktum The average \pt and \xf for each bin are also given.}
  \label{tab:w}
\end{table}

The fact that \lam and \alam polarizations are small near $x_F=0$
compared to the maximal measured value of $\approx -20\%$ at
$x_F\approx 0.5$~\cite{Lundberg:1989hw} is not surprising for at least two
reasons: firstly, previous measurements at positive \xf show that the
magnitude of \lam polarization decreases with $x_F$
\cite{Ramberg:1994tk}, and secondly, in $pp$ collisions, the
polarization must be an antisymmetric function of \xf for symmetry
reasons (i.e. to avoid the ambiguity which would otherwise be 
encountered, since there is no a priori reason to favor the beam proton direction over the target proton direction when defining the production plane at $x_F=0$).  Any non-zero polarization near 
$x_F=0$ in the
present measurement would then necessarily come from either a
difference in \lamalam polarization in interactions with neutrons
compared to protons or nuclear effects.  Previous attempts to measure
nuclear effects in \lamalam polarization show that, in the forward
region, any such effects can only be very weak \cite{Soffer:1999ww}.

\section{Systematic studies}
\label{subsec:systematics}

We discuss separately two categories of possible systematic errors:
those due to possible inaccuracies in the MC detector description
(${\sigma_{acceptance}}$), and those due to the method of
extracting the signal (${\sigma_{method}}$). An additional
rather insignificant contribution to the systematic error is
introduced by the uncertainty of the decay asymmetry parameter 
$\alpha_{\Lambda}$, as obtained from the PDG \cite{Eidelman:2004wy}.
The final estimates for each of these three sources is given in
Table~\ref{tab:settings}.

\begin{table}[h]
\begin{center}
\setlength{\extrarowheight}{2.0pt}
\begin{tabular}{l|c|c|c} 
 \multicolumn{4}{l}{{\quad\quad$\mathbf\Lambda$ }} \\ \hline
 \xf interval& [-0.15; -0.07] & [-0.07; -0.04] & [-0.04; 0.01] \\ \hline \hline
$\sigma_{acceptance}$ & $\pm$0.022&$\pm$0.022&$\pm$0.022 \\ \hline
$\sigma_{method}$&-0.004&\medspace\thinspace0.000&+0.002 \\ \hline
$\sigma_{\alpha}$&$\pm$0.001&\medspace\thinspace0.000&$\pm$0.001 \\ \hline  \hline

\multicolumn{4}{l}{{\quad\quad$\mathbf{\overline{\Lambda}}$ }} \\ \hline
$\sigma_{acceptance}$&$\pm$0.029&$\pm$0.029&$\pm$0.029\\ \hline
$\sigma_{method}$&+0.002&+0.006&-0.003\\ \hline
$\sigma_{\alpha}$&\medspace\thinspace\thinspace\thinspace0.000&$\pm0.001$&\medspace\thinspace0.000\\ \hline

\end{tabular}
\caption{
  The various systematic errors. Note that the contribution from the
  difference in polarization obtained by two different methods is
  one-sided.}
\label{tab:settings}
\end{center}
\end{table}

To establish limits on biases due to an imperfect MC efficiency
determination, we first evaluate the asymmetry in the $cos(\theta_y)$
distributions for various subsamples.  Any asymmetry in
$cos(\theta_y)$ could only be due to detector bias (see
Sec.~\ref{sec:intro}).  The limits are evaluated separately for \lam
and \alam since their decay products traverse rather different parts
of the spectrometer due to bending in the spectrometer magnetic field.
For each of three \xf intervals and for each of the two targets, the
data set is divided into (approximate) halves according to the directions
of the produced \lamalams as seen in the lab frame, e.g.  up/down,
left/right and at various angles in the transverse plane.  For each such pair
of subsamples, the asymmetry difference, a quantity which should be
consistent with zero, is evaluated. 
The largest deviation from zero was found between the up and down 
subsamples. To avoid correlations, we therefore use only
the up/down subsamples as a basis for the evaluation.  The data is
thus divided into a total of 12 subsamples for which the asymmetry is
separately evaluated. The sum, with each term weighted by the inverse
square of its statistical error, is formed and interpreted
as a $\chi^2$ statistic
for 12 degrees of freedom.  The systematic error is estimated by
dividing the obtained number by $1+f^2$ where $f$ is determined by
requiring the resulting $\chi^2$ to correspond to a $50\%$
probability. The systematic error estimate due to possible acceptance
distortions is then the average of the statistical errors in the three
\xf bins multiplied by $f$. The results are given in Table~\ref{tab:settings}.  
Similar results were obtained in a cross check analysis using
$cos(\theta_x)^{\uparrow}-cos(\theta_x)^{\downarrow}$ rather than
$cos(\theta_y)$,  where $cos(\theta_x)^{\uparrow}$ and $cos(\theta_x)^{\downarrow}$ 
refers to \lams propagating in the direction of upper/lower hemisphere 
in the lab frame.\\
As a cross check of the systematic error, the apparent \ks polarization 
was determined using the same method as for $\Lambda/\overline{\Lambda}$.
Since \ks is a pseudo-scalar meson, it cannot be polarized. The result for \ks is that even for $f=0$, the probability for zero polarization exceeds $50\%$.
An additional systematic error, $\sigma_{method}$, could result from 
the fitting procedure used to
extract the number of signal events in each bin.  An alternative to
the fit procedure, namely counting the number of \lamalam candidates
in the signal region of the mass plot and subtracting background, as
estimated from side-bins was checked.

Estimates of the individual contributions to the systematic error 
are shown in Table~\ref{tab:settings}. Of the
three sources considered, the first dominates.
Note also that the first contribution is correlated between the three
\xf bins since the decay products corresponding to the different bins traverse the same detector elements. The second is
proportional to the measured polarization and the third is not
correlated with the first two. To be conservative, the correlations 
are ignored, and the total systematic error (see Table~\ref{tab:combined}) 
is calculated by adding the individual contributions of Table~\ref{tab:settings} in quadrature.

\section{Discussion}
\label{sec:discussion}

The present measurements are performed in three \xf bins which are
integrated over a \pt interval common to all bins.  In contrast, most
previous measurements were performed in relatively small lab-frame
angular apertures, and thus, unlike the present measurement, have
strong correlations between the average \xf and \pt values of the
reported results.  Consequently, a comparison is non-trivial.
Furthermore, a point by point comparison is not possible for two
reasons: very few publications supply all the needed information (the
average \xf and \pt values of the measured points), and, the \xf
region of the present measurement does not overlap with the regions of
previous measurements.  Instead, we compare our results to a
parameterization of measurements given in Ref.~\cite{Lundberg:1989hw},
which describes measurements from four experiments performed at
$400$~GeV proton beam energy with hydrogen and beryllium targets, at
various targeting angles.  Those results cover the \xf range
[0.1;0.5], and are fitted to a simple expression with factorized \xf
and \pt dependences:
\begin{equation}
P_{ext}(x_F,p_{\bot})=(C_1x_F+C_2x_F^3)(1-e^{c_3p_{\bot}^2}).
\label{eq:pxfpt}
\end{equation}
The fitted coefficients are: $C_1=-0.268\pm0.003$,
$C_2=-0.338\pm0.015$ and $C_3=-4.5\pm0.6~\text{(GeV/c)}^{-2}$.
\begin{table}
  \centering
\setlength{\extrarowheight}{2.0pt}
  \begin{tabular}[!ht]{    >{ $}l<{$  }>{ $}c<{$}> {$}c<{$  }>{ $}c<{$}>{ $}c<{$}    }
  \mathbf{\Lambda}  &  <p_{\bot}> & <x_F> & Polarization & P_{ext} \\
&   \text{[GeV/c]}& & & \\ \hline
&  0.82&-0.099&\quad\negthinspace\negthinspace 0.054 \pm 0.019(stat)\pm0.022(sys)& 0.025 \\
&  0.82&-0.055&\quad\negthinspace\negthinspace 0.007 \pm 0.020(stat)\pm0.022(sys)& 0.014 \\ 
&  0.84&-0.020&-0.034 \pm 0.018(stat)\pm0.022(sys)& 0.005 \\ \hline 
\mathbf{\overline{\Lambda}}\\  \hline 
&  0.82&-0.097&-0.024 \pm0.030(stat)\pm 0.029(sys)& \\
&  0.81&-0.054&\quad\negthinspace\negthinspace  0.028\pm0.030(stat)\pm
  0.029(sys) &\\
&  0.83&-0.020&-0.024 \pm0.025(stat)\pm 0.029(sys) & \\ \hline 
  \end{tabular}

\caption[Comparison of measurements obtained in this study with previous measurements]
{Combined \lamalam polarization results for W- and C-target samples in three 
bins of \xfpunktum The average \pt and \xf for each bin are also given. $P_{ext}$ 
is the expected polarization extrapolated from previous measurements.}

\label{tab:combined}
\end{table}

\begin{figure}[h!!!]
 \centering
\epsfig{file=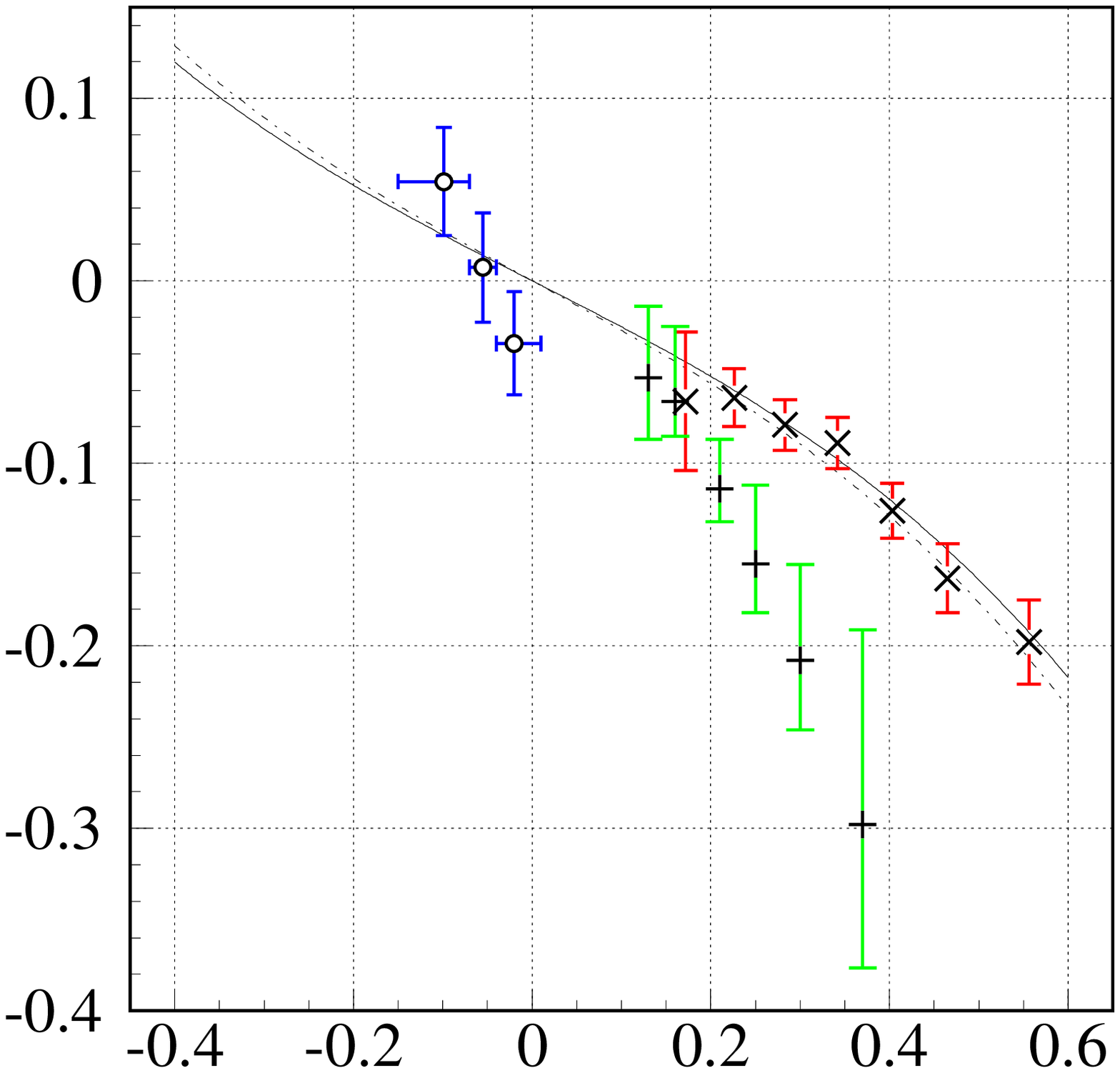,width=0.55\textwidth}
\begin{picture}(0,0)
\put(-68,216){$\times$ Ref.~[2]}
\put(-68,200){$+$ Ref.~[14]}
\put(-68,184){$\circ$ \hb}
\put(-25,3){\Large{\xf}}
\put(-250,153){\rotatebox{90.0}{\Large{Polarization}}}

\end{picture}
\caption[\lam polarization dependence on \xf]{\lam polarization
  dependence on \xfpunktum The curves correspond to Eq.~(\ref{eq:pxfpt})
  \cite{Lundberg:1989hw} for infinite \pt (dashed curve) and for the
  measured \pt spectrum (solid curve).}

\label{fig:results}
\end{figure}

In Ref.~\cite{Heller:1996pq} it is argued that the \lam polarization
dependence on CM energy is weak. Assuming complete energy
independence, the functional form of Eq.~(\ref{eq:pxfpt}) can be
checked against more recent and independent $800$~GeV \lam
measurements~\cite{Ramberg:1994tk}. The comparison is shown in
Fig.~\ref{fig:results}, where the solid curve corresponds to
Eq.~(\ref{eq:pxfpt}) with \pt = 0.77~GeV/c, the \pt equivalent obtained 
by averaging ($1-e^{c_3p_{\bot}^2}$) for the present
measurement and the dashed curve corresponds to Eq.~(\ref{eq:pxfpt}) with
infinite \ptpunktum Except for the lowest \xf point, the measurements of
Ref.~\cite{Ramberg:1994tk} are at larger \pt than the present results,
and should therefore correspond to a curve lying between the two
displayed curves. The data are clearly consistent with the
parameterization. Also shown in Fig.~\ref{fig:results} are the three
\hb measurements, which are also compatible with the extrapolation of
the parameterization to negative \xfpunktum The values of $P_{ext}$
corresponding to the \hb points are given in Table~\ref{tab:combined}.

The third dataset shown in Fig.~\ref{fig:results} are results from NA48~\cite{Fanti:1999}\footnote{In Ref.~\cite{Ramberg:1994tk,Fanti:1999} \xf is defined in the laboratory system. This gives rise to small shifts in the \xf calculation as compared to our definition.}. The NA48 data, albeit taken at a similar $\sqrt{s}$ and
in a similar kinematic regime as the measurements parametrized by Eq.~(\ref{eq:pxfpt}), are not described by this parameterization and are inconsistent with Ref.~\cite{Lundberg:1989hw}.

Previous measurements of \alam polarization include: $0.006\pm0.005$~\cite{Lundberg:1989hw},
$0.014\pm0.027$~\cite{Ramberg:1994tk}, and
$-0.014\pm0.037$~\cite{Fanti:1999}. These numbers are average values
for the specific kinematic ranges covered by each experiment and are
therefore not directly comparable. Nonetheless, all results are
consistent with zero and in agreement with our measurement.

\section{Conclusion}

A measurement of the inclusive \lamalam polarization has been
performed in the \xf range:\newline[-0.15; 0.01] and the \pt interval:
[0.6; 1.2]~GeV/c using \lamalams produced in $pC$ and $pW$
collisions. As the polarization results from the two targets agree within
their statistical uncertainties, we see no evidence of nuclear effects.
The magnitude of the \lam polarization is less than $\approx 6\%$ and
measurements suggest an increase of the polarization with an increase
of $|x_F|$.
When combining the data from the two targets, the largest deviation from zero polarization,~$+0.054 \pm 0.029$, is measured for $x_F\lesssim -0.07$.
Zero polarization is expected at $x_F=0$ in the absence of
nuclear effects.
The \lam polarization measurements are consistent with a
parameterization, $P_{ext}$, of earlier measurements performed at
positive \xfkomma where the polarization is negative.
The \alam polarization measurements are consistent 
with zero.

\section*{Acknowledgments}

We are grateful to the DESY laboratory and to the DESY accelerator group for their strong support since the conception of the \hb experiment. Also, we would like to thank the technical and administrative staff without whom the \hb experiment would not have been possible.

\end{document}

%% file: authors_epjc.tex
I.~Abt$^{23}$,
M.~Adams$^{10}$,
M.~Agari$^{13}$,
H.~Albrecht$^{12}$,
A.~Aleksandrov$^{29}$,
V.~Amaral$^{8}$,
A.~Amorim$^{8}$,
S.~J.~Aplin$^{12}$,
V.~Aushev$^{16}$,
Y.~Bagaturia$^{12,36}$
% ...

\vspace{5mm}
\noindent
% ...
$^{8}${\it LIP Coimbra, P-3004-516 Coimbra,  Portugal~$^{f}$} \\
% ...
$^{10}${\it Institut f\"ur Physik, Universit\"at Dortmund, D-44221 Dortmund, Germany~$^{d}$} \\
% ...
$^{12}${\it DESY, D-22603 Hamburg, Germany} \\
$^{13}${\it Max-Planck-Institut f\"ur Kernphysik, D-69117 Heidelberg, Germany~$^{d}$} \\
% ...
$^{16}${\it Institute for Nuclear Research, Ukrainian Academy of Science, 03680 Kiev, Ukraine~$^{h}$} \\
% ...
$^{23}${\it Max-Planck-Institut f\"ur Physik, Werner-Heisenberg-Institut, D-80805 M\"unchen, Germany~$^{d}$} \\
% ...
$^{29}${\it DESY, D-15738 Zeuthen, Germany} \\
% ...
$^{36}${\it visitor from High Energy Physics Institute, 380086 Tbilisi, Georgia} \\
% ...
$^{d}$ supported by the Bundesministerium f\"ur Bildung und Forschung, FRG, under contract numbers 05-7BU35I, 05-7DO55P, 05-HB1HRA, 05-HB1KHA, 05-HB1PEA, 05-HB1PSA, 05-HB1VHA, 05-HB9HRA, 05-7HD15I, 05-7MP25I, 05-7SI75I \\
% ...
$^{f}$ supported by the Portuguese Funda\c c\~ao para a Ci\^encia e Tecnologia under the program POCTI \\
% ...
$^{h}$ supported by the National Academy of Science and the Ministry of Education and Science of Ukraine \\
% ...
